\documentclass[conference, a4paper]{IEEEtran}

\IEEEoverridecommandlockouts                              
\usepackage{units}
\usepackage{graphicx}
\usepackage{multirow}
\usepackage{amsmath,amssymb,latexsym}
\usepackage{rotating}
\usepackage{lettrine}
\usepackage{textcomp}
\usepackage{hyperref}
\usepackage{xcolor}
\usepackage{url}

\newcommand{\Ntx}{\mathrm{N}_\mathrm{TX}}
\newcommand{\Nrx}{\mathrm{N}_\mathrm{RX}}
\newcommand{\nuex}{$\nu$-exchange}
\newcommand{\consp}{CONSIP}

\usepackage{lipsum}
\usepackage[pscoord]{eso-pic}
\newcommand{\placetextbox}[3]{
  \setbox0=\hbox{#3}
  \AddToShipoutPictureFG*{
    \put(\LenToUnit{#1\paperwidth},\LenToUnit{#2\paperheight}){\vtop{{\null}\makebox[0pt][c]{#3}}}%
  }%
}%

\IEEEoverridecommandlockouts
\title{\consp{}: Consistency Protocol for Hopping Function Exchange and Black listing in TSCH}

\author{Stefano Scanzio, Federico Bitondo, Gianluca Cena, and Adriano Valenzano\\
National Research Council of Italy (CNR--IEIIT), Corso Duca degli Abruzzi 24, I-10129 Torino, Italy\\
Email: stefano.scanzio@ieiit.cnr.it, federico.bitondo@gmail.com,\\ gianluca.cena@ieiit.cnr.it, adriano.valenzano@ieiit.cnr.it\\
}

\begin{document}
\placetextbox{0.5}{1}{This is the author's version of an article that has been published in this journal.}
\placetextbox{0.5}{0.985}{Changes were made to this version by the publisher prior to publication.}
\placetextbox{0.5}{0.97}{The final version of record is available at \href{https://doi.org/10.1109/WFCS53837.2022.9779192}{https://doi.org/10.1109/WFCS53837.2022.9779192}}%
\placetextbox{0.5}{0.05}{Copyright (c) 2022 IEEE. Personal use is permitted.}
\placetextbox{0.5}{0.035}{For any other purposes, permission must be obtained from the IEEE by emailing pubs-permissions@ieee.org.}%

\maketitle
\thispagestyle{empty}
\pagestyle{empty}

\begin{abstract}
The use of white and black listing techniques in Wireless Sensor Networks (WSN), and in particular those which are based on the Time Slotted Channel Hopping (TSCH) operating mode of IEEE 802.15.4, permits to improve reliability and latency by performing transmissions on the best channels. Techniques that operate on a per-link basis are deemed quite effective, but proper operation requires that the two end points involved in the communication agree on the channels to be used for transmission. On the contrary, communication in the network can be prevented, eventually leading, in the worst cases, to the disconnection of part of the nodes.

This paper presents \consp{}, a technique aimed to ensure strict consistency in the information exchanged between the nodes and used to drive communication, 
by preventing a priori the aforementioned problem from occurring. 
Results show a slight increase in energy consumption, due to the use of a backup cell, whereas communication latency does not worsen. 
The effectiveness of \consp{} was assessed by means of an experimental campaign, 
and the only drawback we found is that the backup cell, which is required to be reserved per link,
may limit the number of nodes in dense networks.
\end{abstract}

\section{Introduction}
\label{sec:introduction}
The recent evolution of factories can be viewed as a sort of new, pacific futurism movement\footnote{Futurism was a social and artistic movement, which originated in Italy in the early 20th century.}, 
in which characteristics such as \textit{speed}, \textit{simultaneity}, \textit{dynamism}, and \textit{innovation} 
play a relevant role in next-generation industrial networks and plants. 
On the one hand, the Industry 4.0 \cite{doi:10.1080/00207543.2020.1824085,CANAS2021107379} revolution (and beyond) is an example of this transition, 
where communication networks are becoming more and more heterogeneous in terms of the employed technologies \cite{SCANZIO2021103388} and wireless extensions are playing an increasingly important role for enabling the aforementioned characteristics.
On the other hand, the (Industrial) Internet of Things (IoT/IIoT) paradigm \cite{MALIK2021125, 9381665, 8401919} enables autonomous communication between network devices, 
and is boosting the need of wireless technologies with more specific features, able to meet increasingly demanding requirements.

In this direction, the set of technologies collectively known with the term Wireless Sensor (and Actuator) Networks (WSN/WSAN) is extensively used to collect data and, sometimes, to perform actuations in distributed systems, where devices are typically battery-powered and the amount of data to be exchanged is quite small. 
Among the available solutions, the Deterministic and Synchronous Multichannel Extension (DSME) and the Time Slotted Channel Hopping (TSCH) operating modes of the IEEE 802.15.4 standard \cite{9144691} show interesting features and are becoming quite popular.
In both modes, the transmissions of packets in a data stream, as well as the retransmissions of the same packet, are carried out automatically at the MAC layer on different frequencies/channels,
which consequently makes the quality of the communication link, as seen by network nodes, more stable.

This work focuses on TSCH, and in particular it is related to those techniques, known as black listing and white listing, that are aimed at increasing the quality of communication, by removing the worst channels or by selecting the best channels, respectively. 
For example, if a network node is given the ability to select the best channels to perform its transmissions, it will consequently experience improvements in terms of reliability, latency, and power consumption.
To this end, many black and white listing algorithms were proposed in the scientific literature, and their typical operations can be subdivided into three basic steps:
\begin{enumerate}
    \item \textit{evaluation}: inferring how each channel will likely behave in the near future, 
    typically using statistics collected from the recent past;
    \item \textit{selection}: deciding whether, where, and how to use these channels, that is, planning a selection strategy;
    \item \textit{propagation}: delivering the channel selection strategy to the involved nodes 
    in a consistent way.
\end{enumerate}

The main content of this paper is related to the last point. In particular, it is about ensuring that, at any given time, the sender and receiver nodes on a link use the same channel for communicating. 
In fact, if this property was not guaranteed, communication within the network could be prevented, possibly with severe consequences, which typically consist in the disconnection, sometimes permanent, of one or more nodes. 
To this extent we proposed the \consp{} protocol, which ensures to the involved nodes a reliable exchange of the information about the channels to be used in future transmissions. 
It grounds on the idea to leave two disjoint communications links open during this exchange, 
which virtually operate in parallel, the former based on the old information about channels and the second using the new one. 
Only when both nodes are certain that they have agreed to use the new information, 
the previous communication based on the old channel selection is deactivated.
An extensive experimental campaign based on simulation has been carried out, in order to evaluate the \consp{} functionality from the point of view of a number of statistical indicators, 
and in particular power consumption, which is a very important performance metric for this kind of networks.

In the following, the concepts behind white and black listing are analyzed and described together in Section~\ref{sec:listing}, with an extensive analysis of the state of the art. 
The \consp{} protocol is firstly presented intuitively, and then formally, in Section~\ref{sec:consistency}. Section~\ref{sec:experimental_setup} describes the simulator and the parameters we used in the simulation, including the energy model, 
while results are reported in Section~\ref{sec:results}, 
which precedes Section~\ref{sec:conclusions} that draws some concluding remarks.

\section{Black and White listing}
\label{sec:listing}
The channel hopping technique enables nodes in a wireless network to periodically change the transmission frequency of links in order to mitigate the effects of disturbance and interference on the quality of communication.
The ideas behind channel hopping were proposed more than one decade ago \cite{10.1145/1641876.1641898}, 
and currently they are adopted in several network technologies like WirelessHART \cite{9055118}, Bluetooth \cite{8935348}, and TSCH.
In particular, time in TSCH is divided in \textit{timeslots} of equal length \cite{CENA2020102199, 9187609}, 
while channel hopping \cite{10.1145/1641876.1641898} selects the effective transmission channel 
through a pseudo-random function $\nu$ shared between the transmitter and the receiver. 
In this protocol, traffic is scheduled by reserving timeslots, to permit nodes to be switched off when no data exchange is scheduled for them, consequently saving energy.

A scheduled transmission is identified by a pair of values, 
the \textit{slot offset} ($\mathrm{o}$) and the \textit{channel offset} ($\mathrm{c}$), 
which identify a position in a matrix of dimension $N_\mathrm{slots} \times N_\mathrm{ch}$. 
The schedule repeats over time with period $N_\mathrm{slots}$ slots. 
Typically $N_\mathrm{slots}=101$, and the slot duration is equal to $\unit[20]{ms}$. 
As a consequence, the repetition period of the slotframe (i.e., of the matrix) is $\unit[2.02]{s}$. 
Each slot is identified by a (practically) unique and strictly increasing unsigned integer number $x$, which is known as the Absolute Slot Number (ASN).
In every slot up to $N_\mathrm{ch}$ transmissions can be performed at the same time on distinct links, each one related to a different channel offset $c \in \{1,...,N_\mathrm{ch}\}$.

The shared \textit{hopping function} $\nu$ returns the physical channel $\mathrm{ch}$ used for the actual transmission, and can be expressed as
\begin{equation}
    \operatorname{ch} = \nu(x, c) \triangleq H( (x+c)\ \mathrm{mod}\ N_\mathrm{ch} ),
\end{equation}
where $H(i)$ is the so called \textit{hopping sequence}.
Given an integer value $i \in \{0,...,N_\mathrm{ch}-1\}$, $H(i)$ basically returns the element in position $i$ of an array of dimension $N_\mathrm{ch}$, which encodes the physical channel. 
If protocol parameters are set in such a way that two subsequent transmissions of the same data frame are not spaced by a multiple of $N_\mathrm{ch}$ (as typically happens in real networks),
then they will take place on different physical channels.
The hopping sequence $H(i)$ is usually chosen so that subsequent transmissions take place on channels that are spaced wide enough. 
In this way, retries of the same packet are unlikely to suffer repeatedly from the effect of the same source of interference \cite{7972164}. For instance, in the $\unit[2.4]{GHz}$ band a single Wi-Fi channel can span over multiple IEEE 802.15.4 channels that use the O-QPSK PHY.

Channel hopping has the effect of ``flattening'' network performance, since all channels are used independently of their quality.
In other words, the quality of a link experienced by communicating nodes is about the same as what is found by averaging the quality of channels.
As a consequence, reliability and other performance indicators are less dependent on the quality of any single channel, making communication much more stable.

Two reasonable and intuitive solutions to enhance performance are \textit{black listing} \cite{8647111} and \textit{white listing} \cite{8667348}. 
In the former, channels with the worst performance are excluded, while in the latter only those channels with the better performance are exploited.
Both solutions are based on the idea that some channels can be removed from (or selected for) the hopping sequence in order to maximize the chances that transmission attempts succeed.

\begin{figure*}[t]
	\begin{center}
	\includegraphics[width=2\columnwidth]{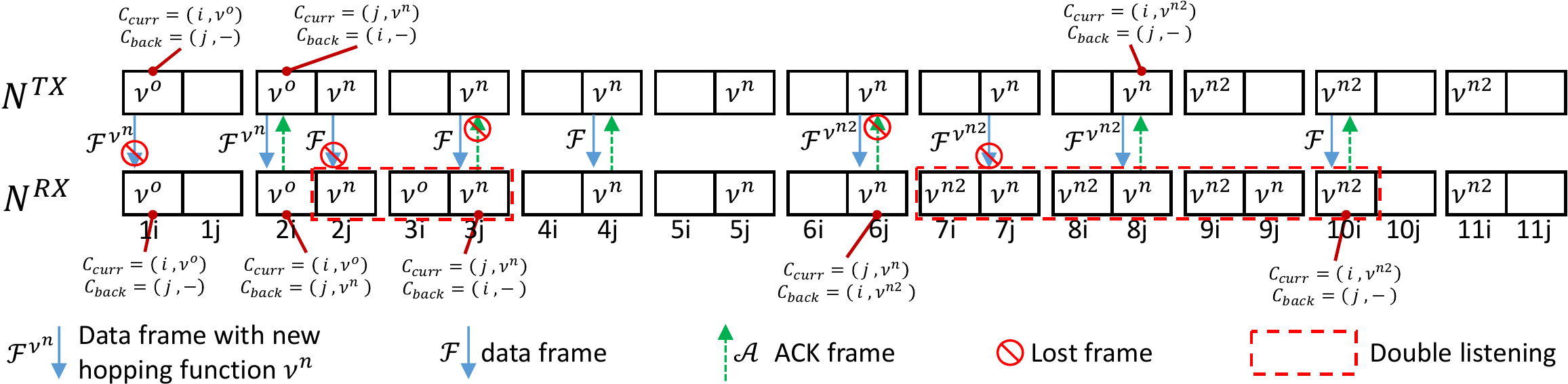}
	\end{center}
	\vspace{-0.2cm}
	\caption{Example of application of the \consp{} protocol.}
	\vspace{-0.2cm}
	\label{fig:consip}
\end{figure*}

Over the past years, a variety of black and white listing techniques have been proposed in the scientific literature. 
A relevant aspect for creating a taxonomy of such techniques is the metric they use to identify a channel as bad or good. 
Methods based on the Received Signal Strength Indicator (RSSI) have been shown to be less accurate than those based on the Packed Delivery Ratio (PDR) \cite{7996980}. 
Less conventional techniques like ETSCH rely on energy detection in idle periods \cite{7366935}, 
while other solutions make use of fuzzy logic \cite{black_listing_fuzzy_logic}. 
It is also possible to use machine learning techniques to predict the evolution of a wireless channel, in terms of the frame delivery probability, given its recent past \cite{https://doi.org/10.1002/itl2.326}.

Channel quality estimation is only one of the aspects to which attention has to be paid in the definition of above techniques. 
Many works recognise that link-based (\textit{local}) listing, in which the set of good/bad channels is selected based on the single link, is better than \textit{global} listing \cite{label}. 
The superiority of link-based approaches is reasonable, since mesh networks are distributed in space, and the characteristics of the wireless spectrum may vary noticeably even in the case of small movements of the nodes \cite{5502548}.

Actually, a node that requires to communicate with more than one neighbor needs to maintain a different $\nu$ function for each neighbor. 
Selecting the correct functions can be done transparently, by using the information related to the configured (non-shared) cells in the slotframe matrix.
An important limitation of link-based techniques is that they cannot be directly used for multicast/broadcast traffic.

Black/white listing techniques also differ on the way channel selection is shared among nodes. 
In particular, every time a technique requires a modification of the function $\nu$ (and/or the hopping sequence $H(i)$) 
used by a pair of nodes, both the transmitter $N_\mathrm{TX}$ and the receiver $N_\mathrm{RX}$ must agree on it.
In the case of inconsistency between the functions $\nu_\mathrm{TX}$ and $\nu_\mathrm{RX}$ used in $N_\mathrm{TX}$ and $N_\mathrm{RX}$, respectively (i.e., when $\nu_\mathrm{TX} \neq \nu_\mathrm{RX}$), communication between the two nodes is prevented because it is no longer guaranteed that the channel they use for transmission and reception in any given slot coincide. 
This circumstance shall be absolutely avoided, because it could lead to the disconnection of a portion of the network.

The ways to distribute $\nu$ to the end points of the link, and in particular to do so consistently, avoiding the possibility that $\nu_\mathrm{TX} \neq \nu_\mathrm{RX}$ for even just a single cell, are the main goals of this work.
Some scientific papers treated this problem in a completely different way and with some assumptions \cite{label,MABO}, for example not considering the chance that acknowledgement (ACK) frames could be lost. 
This simple assumption proved to be mostly untrue on real traffic logs we acquired on OpenMote B devices equipped with the OpenWSN (version REL-1.24.0) operating system. 
In fact, our measurements showed that the loss probability for ACK frames in our experimental testbed is not negligible at all,
and amounts to about $8\%$ in benign environmental conditions. 
In the same conditions, the loss probability for data frames was $12.6\%$\footnote{The experimental data on which such values were computed are included in the file \texttt{default-101-16-15days.dat}, which is downloadable from \href{https://dx.doi.org/10.21227/fg62-bp39}{https://dx.doi.org/10.21227/fg62-bp39}.}.

Instead, the main idea in \cite{electronics11030304} is to leave the hopping function unmodified, 
but to lower the usage of cells associated with bad quality channels, which are exploited with a certain probability lower than one. 
Doing so prevents any problems due to inconsistency between the views of $\Ntx$ and $\Nrx$, 
but at the same time limits the achievable performance.
In fact, while this technique is suitable for reducing power consumption, it also causes an increase of latency.

\section{Consistency Protocol}
\label{sec:consistency}
Communication between a sender node $\Ntx$ and a receiver node $\Nrx$ in the slot characterized by ASN equal to $x$ is only possible if both nodes agree on the frequency (physical channel) to be used to send and receive data on air, respectively. 
In the case the hopping functions $\nu$ on the two sides of a link were unaligned, the network would quickly lose connectivity of some nodes, which become unreachable.

The process of exchanging a hopping function between the end points of a link starts when $\Ntx$ generates a new $\nu^n$ and finishes when $\nu^n$ is effectively in use, 
at which point the cells referring to the previous hopping function $\nu^o$ are switched off in both $\Ntx$ and $\Nrx$. 
In the following, this process will be denoted, for brevity, with the term \textit{\nuex}. 
Instead, \consp{} is the protocol we are proposing in this work to enable a \nuex{} to be performed in a consistent way.

The main idea behind \consp{} is to have, for any cell $C_\mathrm{curr}$ reserved for the communication between $\Ntx$ and $\Nrx$, a backup cell $C_\mathrm{back}$ that is used only for the time strictly needed to perform a \nuex. 
After that, the cell $C_\mathrm{back}$  becomes $C_\mathrm{curr}$, i.e., the two cells reverse their role by means of a \texttt{swap()} operation.
These two cells are scheduled in distinct slot offsets in the slotframe matrix. 
As a consequence, \consp{} does not require any modification to the hardware of nodes. 
In particular, it can be implemented in conventional devices provided with a single communication interface/antenna. 
Although there are no other constraints on the position of $C_\mathrm{back}$ within the slotframe, a reasonable choice is to interleave $C_\mathrm{curr}$ and $C_\mathrm{back}$ cells so that they are evenly spaced, e.g., when $N_{\mathrm{slots}}=101$ they could be located at slot offset $i$ and $(i+50)\!\mod\!101$. 
Both $C_\mathrm{curr}$ and $C_\mathrm{back}$ are identified by a 2-tuple, composed of a slot offset and a hopping function. 
Let $i$ and $j$ be the slot offsets assigned to the two above cells, respectively.
Then, $C_\mathrm{curr}=(i, \nu^o)$ means that the current cell in use $C_\mathrm{curr}$ is assigned to slot offset $i$ and that hopping function $\nu^o$ is used for transmission.
Instead, $C_\mathrm{back}=(j, -)$ means that $C_\mathrm{back}$ is assigned to slot offset $j$ and it cannot be used because it is not currently mapped to any hopping function.

\begin{figure*}[t]
	\begin{center}
	\includegraphics[width=2\columnwidth]{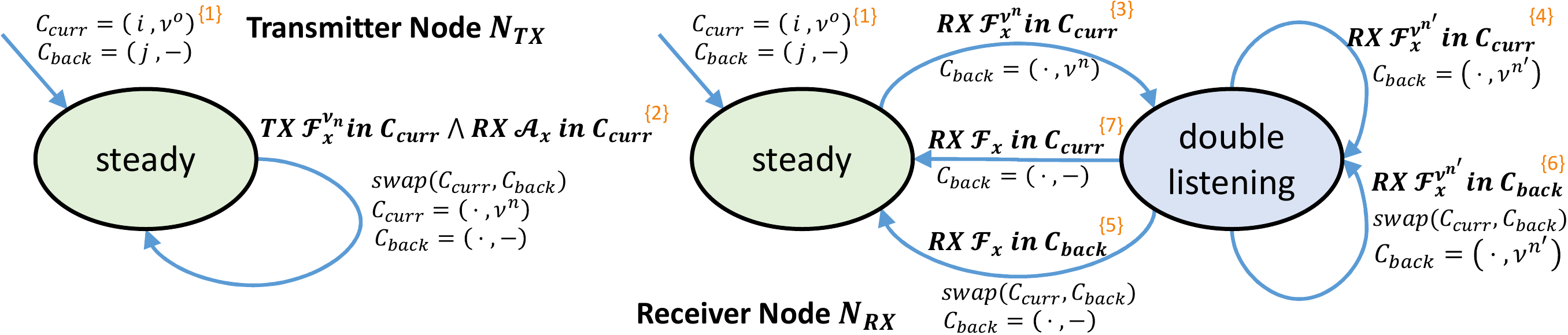}
	\end{center}
	\vspace{-0.2cm}
	\caption{State machines of $\Ntx$ and $\Nrx$.}
	\vspace{-0.2cm}
	\label{fig:machines}
\end{figure*}

\subsection{Simple example about protocol operation}
In Fig.~\ref{fig:consip}, an example of the \consp{} operation is sketched. 
In slot $1i$ (i.e., slot $i$ in slotframe $1$) a frame $\mathcal{F}_x^{\nu^n}$ is sent to perform a \nuex{}.
This frame is just a conventional data frame $\mathcal{F}_x$ that also contains the new hopping function $\nu^n$ estimated by $\Ntx$. 
The hopping function used to determine the physical channel for every slot, on either the transmitter (upper diagram) or the receiver side (lower diagram), is specified inside the box representing the slot itself. 
In the example, the first frame is lost but the retransmission of $\mathcal{F}_x^{\nu^n}$ in cell $2i$ correctly arrives to destination.
For $\Nrx$, starting from slot $2j$ the cell $C_\mathrm{back}=(j,\nu^n)$ is activated, and the node enters the \textit{double listening} state in which it listens on both cells $C_\mathrm{curr}$ and $C_\mathrm{back}$, using for them the hopping functions $\nu^o$ and $\nu^n$, respectively. 
Since the frame $\mathcal{F}_x^{\nu^n}$ sent in cell $2i$ is confirmed by the related ACK frame $\mathcal{A}_x$, $\Ntx$ starts using exclusively the new hopping function $\nu^n$.
In particular, it performs a \textit{swap} between the two cells by means of the \texttt{swap()} function, after which $C_\mathrm{curr}=(j,\nu^n)$ and $C_\mathrm{back}=(i,-)$, i.e., the backup cell is deactivated.
On $\Nrx$, the double listening state persists until it receives a frame (typically in $C_{back}$, but sometimes in $C_{curr}$) whose channel is selected using the new hopping function $\nu^n$. 
Only in this case, $\Nrx$ can be sure that also $\Ntx$ has switched to $\nu^n$, and is consequently using $C_{back}$ as current transmission channel.
At this point $\Nrx$ sets $C_\mathrm{curr}=(j,\nu^n)$ and $C_\mathrm{back}=(i,-)$. 
In the example of Fig.~\ref{fig:consip} this happens in slot $3j$, which becomes the current channel.

In the new \nuex{} with hopping function $\nu^{n2}$ in slot $6j$, the $\mathcal{A}_x$ frame used to confirm $\mathcal{F}_x^{\nu^{2n}}$ is lost. 
In this case, node $\Ntx$ continues to transmit in $C_{curr}=(j,\nu^{n})$, and only when $\mathcal{F}_x^{\nu^{2n}}$ is followed by the correct reception of $\mathcal{A}_x$, 
which happens in slot $8j$, node $\Ntx$ can start transmitting using the new hopping sequence. 
It is worth remarking that, the double listening period in which $\Nrx$ is active on both cells with both the old and new hopping sequences is mandatory because, after the reception of $\mathcal{F}_x^{\nu^{n2}}$ using $\nu^n$, from the $\Nrx$ viewpoint it is not possible to know if $\Ntx$ will use $\nu^{n}$ or $\nu^{n2}$ for the next transmission. 
In fact, if the acknowledgement frame $\mathcal{A}_x$  related to $\mathcal{F}_x^{\nu^{n2}}$ is correctly received by $\Ntx$, it can safely assume that $\Nrx$ received the new hopping function $\nu^{n2}$ and hence it can start using it as the current cell $C_\mathrm{curr}=(i,\nu^{n2})$, 
otherwise it must assume that $\nu^{n2}$ was not received. 
This mismatch of viewpoints between $\Ntx$ and $\Nrx$ may occur in any communications network, 
and it is accentuated in wireless networks where the probability of losses is not negligible at all.

Only when the frame $\mathcal{F}_x$ is actually received in the cell related to the new hopping function $\nu^{n2}$ (in slot $10i$ in the example), node $\Nrx$ can start using $C_\mathrm{curr}=(i,\nu^{n2})$ as the current cell, and it can disable the backup cell $C_\mathrm{back}=(j,-)$.

\subsection{Protocol description through FSMs}
To better formalize the \consp{} protocol, two finite state machines (FSMs)  are presented in Fig.~\ref{fig:machines} for $\Ntx$ and $\Nrx$. 
In particular, both FSMs start from the \textit{steady} initial states with $C_\mathrm{curr}=(i,\nu^{o})$ and $C_\mathrm{back}=(j,-)$ (see the arc labeled \{1\} in the figure).
These initial \textit{steady} states represent the normal operating condition of TSCH,
in which only one cell is active and there are no ongoing updates of $\nu$.

Regarding arc {\{2\}} of the $\Ntx$ FSM, each time a frame delivering a new hopping function $\mathcal{F}_x^{\nu^n}$ is acknowledged in $C_\mathrm{curr}$, the node $\Ntx$ can set the cell $C_\mathrm{curr}$  to $\nu^n$. 
For doing so, the node swaps the two cells (i.e., it invokes \texttt{swap(}$C_\mathrm{curr}, C_\mathrm{back}$\texttt{)}), and then it sets the hopping function $\nu^n$ in the current cell and disables the hopping function, and consequently the ability to transmit, in the backup cell. 
This is performed by means of the two operations $C_\mathrm{curr}=(\cdot,\nu^{n})$ and $C_\mathrm{back}=(\cdot,-)$, where the symbol ``$\cdot$'' means that the slot offset is not changed.

Regarding the FSM of $\Nrx$, it consists of two states. 
Each time a frame containing a new hopping function $\nu^n$ is received in $C_\mathrm{curr}$, 
the backup cell is activated and the FSM enters the \textit{double listening} state through arc \{3\}, in which $\Nrx$ receives on both cells, consequently increasing its power consumption.
The arc labelled \{4\} either corresponds to the $\Ntx$'s
intention to change on the fly the previously transferred hopping function $\nu^{n}$ with a new hopping function $\nu^{n'}$ or, more typically, it is due to a retransmission of the previous frames for which the related ACK did not arrive to destination. 
In the latter case $\nu^{n'}=\nu^{n}$.

The arrival of a frame in the backup cell, as for arc \{5\}, confirms to $\Nrx$ that the transmitter is correctly using the new hopping function.
As a consequence, the receiving node can start using the backup cell as current cell, by invoking \texttt{swap(}$C_\mathrm{curr}, C_\mathrm{back}$\texttt{)}, after which it can disable the backup cell $C_\mathrm{back}=(\cdot,-)$.

A frame $\mathcal{F}_x^{\nu^{n'}}$ delivering a new hopping function $\nu^{n'}$ that is received in the backup channel, as depicted for arc \{6\}, has simultaneously two consequences. 
The first is to confirm to $\Nrx$ that $\Ntx$ is using the new hopping function $\nu^{n}$, 
and the second to communicate that $\Ntx$ wants to perform a new \nuex{} using $\nu^{n'}$. 
For this reason the use of the \texttt{swap()} function is required. 

Finally, arc \{7\} accounts for the unlikely case that $\Ntx$ decides to abort the \nuex{}, 
and it continues transmitting normal frames in $C_\mathrm{curr}$.
For instance, this may happen if the sender node detects that the quality of channels has changed
and the current hopping function $\nu^{o}$ is again the best one.

\section{Experimental Setup}
\label{sec:experimental_setup}
A discrete event simulator named TSCH-predictor, which was developed within the \texttt{SimPy} framework, was used to evaluate the effectiveness of \consp{}. 
Differently from other more common simulators such as TSCH-Sim \cite{s20195663} and 6TiSCH \cite{6TiSCH_sim}, TSCH-predictor has the advantage to be noticeably simpler, and consequently it permits to easily implement and evaluate new algorithms and techniques based on TSCH.
TSCH-predictor was profitably used in other research works like \cite{10.1007/978-3-030-61746-2_11, PRIL_etfa2021}, and more information about its features can be found in \cite{PRIL_etfa2021}.

\begin{table*}[th]
  \caption{Power consumption and latency vs. different exchange periods of $\nu$.}
  \label{tab:power}
  \small
  \begin{center}
    \tabcolsep=0.15cm	
    \renewcommand{\arraystretch}{1.1}
    \begin{tabular}{l|cc||c|ccc|cc||ccccc}
    \multicolumn{3}{c||}{} & \multicolumn{6}{c||}{Power consumption} & \multicolumn{5}{c}{Latency} \\
    Listing & $T_{\mathrm{update}}$ & $\#_\nu$ & $P_{\mathrm{tx/tot}}^{\Ntx}$ & $P_{\mathrm{rx}}^{\Nrx}$ & $P_{\mathrm{listen}}^{\Nrx}$ & $P_{\mathrm{tot}}^{\Nrx}$ & \multicolumn{2}{c||}{$P_{\mathrm{tot}}$} & $\mu_d$ & $\sigma_d$ & $d_{p99}$ & $d_{p99.9}$ & $d_{max}$ \\
    & $[\unit[]{min}]$ & & $[\unit[]{\mu W}]$ & \multicolumn{3}{c|}{$[\unit[]{\mu W}]$} & $[\unit[]{\mu W}]$ & $[\%]$ & \multicolumn{5}{c}{$[\unit[]{s}]$} \\
    \hline \hline
    
    Disabled   & - & - & 8.622 & 9.823 & 62.596 & 72.419 & 81.041 & - & 1.311 & 1.006 & 4.900 & 7.200 & 9.460 \\
    \hline
        \multirow{6}{*}{\begin{sideways}\parbox{1.8cm}{Enabled\\$T_{\mathrm{app}}=\unit[30]{s}$}\end{sideways}} 
    & 7.5 & 0.94 & 8.711 & 9.880 & 67.151 & 77.031 & 85.742 & +5.80\% & 1.311 & 1.006 & 4.900 & 7.200 & 9.480 \\
    & 15  & 1.88 & 8.667 & 9.851 & 64.873 & 74.725 & 83.391 & +2.90\% & 1.311 & 1.006 & 4.900 & 7.200 & 9.460 \\
    & 30  & 3.75 & 8.645 & 9.837 & 63.735 & 73.572 & 82.216 & +1.45\% & 1.311 & 1.006 & 4.900 & 7.200 & 9.440 \\
    & 60  & 7.5 & 8.633 & 9.830 & 63.166 & 72.995 & 81.629 & +0.73\% & 1.311 & 1.006 & 4.900 & 7.220 & 9.480 \\
    & 120 & 15 & 8.628 & 9.826 & 62.881 & 72.707 & 81.335 & +0.36\% & 1.311 & 1.006 & 4.900 & 7.200 & 9.460 \\
    & 240 & 30 & 8.625 & 9.824 & 62.739 & 72.563 & 81.188 & +0.18\% & 1.311 & 1.006 & 4.900 & 7.220 & 9.480 \\
    \hline \hline
    
        Disabled   & - & - & 51.732 & 58.932 & 33.995 & 92.927 & 144.659 & - & 1.384 & 1.091 & 5.340 & 7.880 & 21.660 \\
    \hline
        \multirow{6}{*}{\begin{sideways}\parbox{1.8cm}{Enabled\\$T_{\mathrm{app}}=\unit[5]{s}$}\end{sideways}} 
    & 7.5 & 5.63 & 51.820 & 58.990 & 34.748 & 93.737 & 145.557 & +0.621\% & 1.383 & 1.090 & 5.340 & 7.880 & 20.640 \\
    & 15  & 11.25& 51.776 & 58.961 & 34.371 & 93.332 & 145.108 & +0.311\% & 1.383 & 1.091 & 5.340 & 7.880 & 21.660 \\
    & 30  & 22.5 & 51.754 & 58.946 & 34.183 & 93.130 & 144.883 & +0.155\% & 1.384 & 1.091 & 5.360 & 7.880 & 21.660 \\
    & 60  & 45   & 51.743 & 58.939 & 34.089 & 93.028 & 144.771 & +0.078\% & 1.384 & 1.091 & 5.340 & 7.880 & 20.640 \\
    & 120 & 90   & 51.737 & 58.936 & 34.042 & 92.978 & 144.715 & +0.039\% & 1.384 & 1.091 & 5.340 & 7.880 & 21.660 \\
    & 240 & 180  & 51.734 & 58.934 & 34.019 & 92.952 & 144.687 & +0.019\% & 1.384 & 1.091 & 5.360 & 7.880 & 21.660 \\
    \hline \hline
  
        \end{tabular}
        \vspace{-0.0cm}
  \end{center}
\end{table*}

The main settings used in all the experimental campaigns were selected as follows:
$N_\mathrm{slots}=101$, slot duration $\unit[20]{ms}$, $N_\mathrm{ch}=16$, frame loss probability $\epsilon_f=12.6\%$, and ACK loss probability $\epsilon_a=8.0\%$.
In the experiments, both probabilities $\epsilon_f$ and $\epsilon_a$ were left constant over time.
This is not a big limitation, as the sensitivity of \consp{} with respect to channel errors is not the main focus of this work.
However, such analysis may be of interest, and is left as future work.
As specified, these two values were derived from a real experimental testbed deployed in our lab. 
Finally, the duration of each experiment was set to $\unit[10]{years}$, which ensures for results good statistical significance.

The size of each packet sent in the simulation is $L_\mathrm{tot}=L_\mathrm{header}+L_\mathrm{payload}+L_\mathrm{IE}$, where $L_\mathrm{header}=
\unit[29]{B}$ is the overall size of both the PHY and  MAC headers, $L_\mathrm{payload}=\unit[30]{B}$ refers to the payload (we chose a relatively small size for it, as happens in typical industrial networks and WSANs), 
and $L_\mathrm{IE}$ concerns the \textit{information element} (IE). 
In particular, in this work the IE is used to encode and transfer $\nu$. 
The IE, whose size is $L_\mathrm{IE}=L_\mathrm{IE_h}+L_\mathrm{IE_p}$, is a specific configurable attribute that the IEEE 802.15.4 standards permits to attach to a frame, and consists of an IE header $L_\mathrm{IE_h}=\unit[2]{B}$ and an IE payload $L_\mathrm{IE_p}$ that is a configuration parameter and must have enough room to store $\nu$. 
In the following, the value $L_\mathrm{IE_p}$ was fixed to $\unit[16]{B}$ unless explicitly stated.
The exact way the $\nu$ function is encoded depends on the specific implementation of the black/white listing technique, and is outside the scope of this work.

In this new version of the simulator we exploited the energy model described in \cite{9483595}. 
In particular, the energy to transmit a data frame is $E_{\mathrm{tx}}=E_{\mathrm{tx_0}}+e_{\mathrm{tx}} \cdot L_\mathrm{\mathrm{tot}}$, 
and the energy to receive a data frame is $E_{\mathrm{rx}}=E_{\mathrm{rx_0}}+e_{\mathrm{rx}} \cdot L_\mathrm{\mathrm{tot}}$, where $E_{\mathrm{tx_0}}=\unit[7]{\mu J}$, $e_{\mathrm{tx}}=\unit[2]{\mu J/B}$, $E_{\mathrm{rx_0}}=\unit[65]{\mu J}$, and $e_{\mathrm{rx}}=\unit[1.3]{\mu J/B}$. 
Instead, the energy to transmit an ACK frame (whose size is $\unit[33]{B}$) is $E_{\mathrm{tx}}^{\mathrm{ACK}}=\unit[106]{\mu J}$, 
the energy to receive it is $E_{\mathrm{rx}}^{\mathrm{ACK}}=\unit[79]{\mu J}$, 
and the energy spent for \textit{idle listening}, 
when the receiver switches its interface on without receiving any data,
is $E_{\mathrm{listen}}=\unit[138]{\mu J}$.

\section{Results}
\label{sec:results}
The effectiveness of \consp{} was assessed through an extensive experimental campaign aimed at analyzing it from the point of view of two relevant key performance indicators, namely, power consumption and \nuex{} latency. 
There is no need to analyze it also from the point of view of reliability, because in \consp{}  every packet is guaranteed the same number of retries (upon errors) as the unmodified TSCH.

\subsection{Power consumption}
The first set of experiments is targeted at analyzing the amount of energy used by \consp{} when a \nuex \ is performed cyclically with a period equal to $T_{\mathrm{update}}$.
Several values are chosen for this parameter, in the range from some minutes to a few hours.
Two separate periods were set for the traffic flow between $\Ntx$ and $\Nrx$, that is, 
$T_{\mathrm{app}}=\unit[30]{s}$ and $T_{\mathrm{app}}=\unit[5]{s}$. 
The former value mimics a reasonable generation period for sensors located in leaf nodes. 
Instead, the latter value (i.e., $T_{\mathrm{app}}=\unit[5]{s}$) models the links between nodes near the root, in which the aggregation of the traffic generated from the lower layers of the topology increases the amount of cells that are actually used for transmission. 
Selecting periodic transmission patterns does not limit in any way the validity of the proposed method.

Table~\ref{tab:power} compares network behavior for different values of $T_{\mathrm{update}}$ (and $T_{\mathrm{app}}$) with respect to the case when \consp{} is disabled.
Comparison with conventional TSCH operation (i.e., when \consp{} is disabled) is quite relevant, 
because it permits to statistically detect possible drawbacks of this method. 
We did not find in the scientific literature any methods similar to \consp{} to perform a meaningful comparison.

Starting from $T_{\mathrm{update}}$ and $T_{\mathrm{app}}$, the number of samples for each channel that can be exploited to compute $\nu^n$ is $\#_\nu=\frac{T_{\mathrm{update}} \cdot 60}{T_{\mathrm{app}}\cdot 16}$ (the update time is expressed in minutes).
This value just provides an indication about the amount of new information that can be exploited, on average, by a black/white listing algorithm to compute $\nu^n$.
However, it is not related to the way this computation is actually performed, and not even to the quality of the channel.
More important, it is not relevant to \consp{}. 
The value of $\#_\nu$ is reported in the related column of the table.
When $T_{\mathrm{app}}=\unit[30]{s}$, in order to have at least one sample per channel on every update of the hopping function (that is, $\#_\nu \geq 1$) the update period $T_{\mathrm{update}}$ must be set to a value greater than $\unit[8]{min}$.

Regarding $\Ntx$, all the energy related to communication is spent for data transmission. 
The reason why $P_{\mathrm{tx/tot}}^{\Ntx}$ is inversely proportional to $T_{\mathrm{update}}$ is that, every time a \nuex{} is triggered, the size of the packet to be transmitted is increased by $L_\mathrm{IE}$ bytes.

Regarding power consumption on $\Nrx$, there is a sensible increase in the energy $P_{\mathrm{listen}}^{\Nrx}$ that is wasted because of idle listening. 
This is due to the fact that when $\Nrx$ is in the \textit{double listening} state, 
it enables its receiving interface in both cell $C_{\mathrm{curr}}$ and $C_{\mathrm{back}}$, 
but one of the two cells remains unused because the sender node $\Ntx$ transmits only in one cell, either $C_{\mathrm{curr}}$ or $C_{\mathrm{back}}$. 
In particular, the growth in both the total power consumption $P_{\mathrm{tot}}$ and the power consumption on $\Nrx$ that is observed when \consp{} is employed mostly depends on the increase of $P_{\mathrm{listen}}$, i.e., to the higher amount of idle listening because of the aforementioned double listening.

Analyzing the case $T_{\mathrm{app}}=\unit[30]{s}$, when $\#_\nu\simeq 1$ a new $\nu^n$ can be obtained exploiting, on average, about one additional sample for each channel. 
This can be likely considered a worst condition from the point of view of energy, 
and the relative increase of the total power consumption ($P_{\mathrm{tot}}$) is equal to $\unit[+5.80]{\%}$, which is a reasonable value for many application contexts. 
When the \nuex{} is performed at a slower pace (i.e., every $T_{\mathrm{update}} = \unit[30]{min}$ and $\unit[60]{min}$, which means $\#_\nu=3.75$ and $\#_\nu=7.5$) the relative increase in terms of total power consumption is almost negligible and equal to $\unit[+1.45]{\%}$ and $\unit[+0.73]{\%}$, respectively. 
This confirms that the main drawback of \consp{} is not the additional energy consumption, 
but the need to allocate twice as much the number of cells for each link if compared with a scheduling strategy without \consp{}. 
 The results with $T_{\mathrm{app}}=\unit[5]{s}$, in which case the maximum relative increase in the total power consumption is only $\unit[+0.621]{\%}$, further corroborate this conclusion. 
However, this limitation is typically problematic only for a small subset of network topologies, 
characterized by a larger number of nodes and a high density.

The last set of columns in Table~\ref{tab:power} reports some performance indicators related to latency. 
They show that the influence of \consp{} on latency is irrelevant. 
The only statistical indices that are not the same for all the experimental conditions are percentiles ($d_{p99.9}$ for $T_{\mathrm{app}}=\unit[30]{s}$, and $d_{p99}$ for $T_{\mathrm{app}}=\unit[5]{s}$) and the maximum value ($d_{max}$), but they are anyway very similar. 
This behaviour is somehow expected, since high-order percentiles and the maximum converge to their real values much more slowly than other statistical indices such as the mean value and lower-order percentiles.

\begin{table}[th]
  \caption{Power consumption with $T_{\mathrm{app}}=\unit[30]{s}$ and $T\_{\mathrm{update}}=\unit[30]{min}$ vs. different sizes of $\nu$.}
  \label{tab:power2}
  \small
  \begin{center}
    \tabcolsep=0.15cm	
    \renewcommand{\arraystretch}{1.1}
    \begin{tabular}{c|c|ccc|cc}
    $L_\mathrm{IE_p}$ & $P_{\mathrm{tx/tot}}^{\Ntx}$ & $P_{\mathrm{rx}}^{\Nrx}$ & $P_{\mathrm{listen}}^{\Nrx}$ & $P_{\mathrm{tot}}^{\Nrx}$ & \multicolumn{2}{c}{$P_{\mathrm{tot}}$} \\
    $[\unit[]{B}]$ & $[\unit[]{\mu W}]$ & \multicolumn{3}{c|}{$[\unit[]{\mu W}]$} & $[\unit[]{\mu W}]$ & $[\%]$ \\
    \hline \hline
    16 & 8.645 & 9.837 & 63.735 & 73.572 & 82.216 & +1.450\% \\
    14 & 8.642 & 9.835 & 63.735 & 73.570 & 82.212 & +1.445\% \\
    12 & 8.639 & 9.833 & 63.735 & 73.568 & 82.207 & +1.439\% \\
    10 & 8.636 & 9.832 & 63.735 & 73.567 & 82.203 & +1.433\% \\
    8  & 8.633 & 9.830 & 63.735 & 73.565 & 82.198 & +1.428\% \\

    \hline \hline
        \end{tabular}
        \vspace{-0.3cm}
  \end{center}
\end{table}

Table~\ref{tab:power2} shows the results of another experimental campaign aimed at analyzing the effect on power consumption of the size $L_\mathrm{IE_p}$ of the encoding of $\nu$. 
The values $T_{\mathrm{app}}=\unit[30]{s}$ and $T_{\mathrm{update}}=\unit[30]{min}$ are representative of typical operating conditions, and therefore they have been left unmodified. 
As highlighted in the rightmost columns of the result table, 
the relative increase of $P_{\mathrm{tot}}$ ranges from $\unit[+1.428]{\%}$ (when $L_\mathrm{IE_p}=\unit[8]{B}$) 
to $\unit[+1.450]{\%}$ (when $L_\mathrm{IE_p}=\unit[16]{B}$), which means that advanced optimizations on the way $\nu$ is encoded, that lead to a further reduction of $L_\mathrm{IE_p}$, 
only lead to insignificant improvements from the point of view of energy consumption. 
However, its reduction is important because it permits to increase the room available for the payload in the frame.

An additional experimental campaign was carried out, again with $T_{\mathrm{app}}=\unit[30]{s}$ and $T_{\mathrm{update}}=\unit[30]{min}$, to analyze the effect of the placement of the two cells, $C_{\mathrm{curr}}$ and $C_{\mathrm{back}}$.
Two cases were considered, where the cells were equally spaced (in the slot offsets $1$ and $51$, respectively) and contiguous (in the slot offsets $1$ and $2$, respectively). 
As expected, we verified that these is no influence on power consumption, and the influence on latency was irrelevant. 
For instance, the average latency was the same in the two cases, while standard deviation passes from $\sigma_d=\unit[1005.81]{ms}$ in the case of equally spaced cells to $\sigma_d=\unit[1006.07]{ms}$ in the case of contiguous cells.

\subsection{Update latency}

\begin{figure}[t]
	\begin{center}
	\includegraphics[width=0.9\columnwidth]{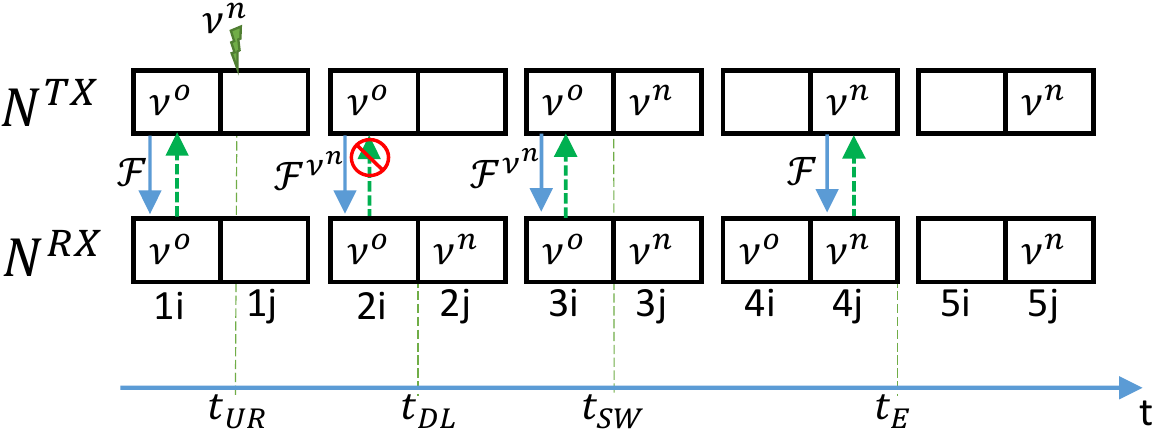}
	\end{center}
	\vspace{-0.2cm}
	\caption{Main temporal events involved in a \nuex{} in \consp{}.}
	\vspace{-0.2cm}
	\label{fig:times}
\end{figure}

Another important performance indicator is the time to complete a \nuex{}. 
In fact, at the end of the whole process, when $\Nrx$ returns in the \textit{steady} state, 
only one of the two cells is effectively used to transmit, and from that point on the system behaves as standard TSCH, with the only exception that $C_{\mathrm{back}}$ remains reserved for future exchanges, although not in use.

Referring to Fig.~\ref{fig:times}, to analyze timings four points in time were identified, 
in correspondence to the main events that make up a \nuex{}. 
Timestamps on these specific events are acquired with the resolution of the ASN (one slot time),
which in this experimental campaign corresponds to $\unit[20]{ms}$. 
In particular, they are:
\begin{enumerate}
\item $t_{\mathrm{UR}}$ (\textit{update request}) is the time when a \nuex{} is started.
\item $t_{\mathrm{DL}}$ (\textit{double listening}) represents the time when a new hopping function $\nu^n$ is received by $\Nrx$. 
At this time the receiver enters the \textit{double listening} state, 
in which it hears from both cells $C_{\mathrm{curr}}$ and $C_{\mathrm{back}}$.
\item $t_{\mathrm{SW}}$ (\textit{swap}) is the time when $\Ntx$ starts transmitting using the new hopping function $\nu^n$.
\item $t_{\mathrm{E}}$ (\textit{end}) is the time when $\Nrx$ starts receiving only using the new hopping function $\nu^n$, 
consequently exiting the \textit{double listening} state. This ends the whole \nuex{} process.
\end{enumerate}

Starting from these four timestamps, we analyzed three main kinds of latency, namely:
\begin{itemize}
\item $d_{\mathrm{SW}}=t_{\mathrm{SW}}-t_{\mathrm{UR}}$ (\textit{swap latency}) is the time elapsing from the beginning of a \nuex{} to the time when the new hopping function $\nu^n$ is actually used to transmit data from $\Ntx$ to $\Nrx$.
\item $d_{\mathrm{DL}}=t_{\mathrm{E}}-t_{\mathrm{DL}}$ (\textit{double listening latency}) is the time interval for which $\Nrx$ remains in the \textit{double listening} state. This interval is characterized by a higher amount of energy consumption.
\item $d_{\mathrm{tot}}=t_{\mathrm{E}}-t_{\mathrm{UR}}$ (\textit{total latency}) is the time needed to complete the whole \nuex{} process, starting from the request and up to the update of the hopping function.
\end{itemize}

\begin{table}[b]
  \caption{Latency related to a \nuex{} in \consp{}.}
  \label{tab:latencies}
  \footnotesize
  \begin{center}
    \tabcolsep=0.13cm	
    \renewcommand{\arraystretch}{1.1}
    \begin{tabular}{l|cccccc}
    Latency & $\mu_d$ & $\sigma_d$ & $d_{\mathrm{min}}$ & $d_{\mathrm{p99}}$ & $d_{\mathrm{p99.9}}$ & $d_{\mathrm{max}}$ \\
    & \multicolumn{6}{c}{$[\unit[]{s}]$} \\
    \hline \hline
    $d_{\mathrm{SW}}=t_{\mathrm{SW}}-t_{\mathrm{UR}}$ & 1.491 & 1.256 & 0.0 & 5.880 & 9.080 & 14.100 \\
    $d_{\mathrm{DL}}=t_{\mathrm{E}}-t_{\mathrm{DL}}$ & 30.005 & 1.511 & 19.180 & 33.340 & 35.360 & 41.420 \\
    $d_{\mathrm{tot}}=t_{\mathrm{E}}-t_{\mathrm{UR}}$ & 31.294 & 1.009 & 30.000 & 34.900 & 37.200 & 41.660 \\

 \end{tabular}
  \end{center}
\end{table}

A further experiment was carried out where $T_{\mathrm{app}}=\unit[30]{s}$ and $T_{\mathrm{update}}=\unit[30]{min}$.
The main statistic indicators we obtained for the above latencies are reported in Table~\ref{tab:latencies}. 
Regarding the swap latency $d_{\mathrm{SW}}$, its average is $\unit[1.491]{s}$.
\begin{figure}[t]
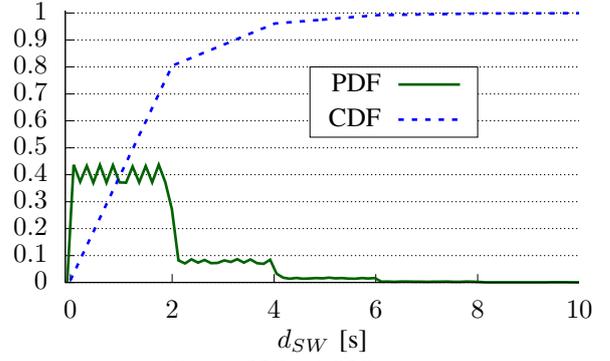

	\begin{center}
	\include{latency}
	\end{center}
	\vspace{-0.4cm}
	\caption{PDF and CDF of $d_{\mathrm{SW}}$.}
	\vspace{-0.2cm}
	\label{fig:pdf_times}
\end{figure}
This value can be explained by analyzing the plot of the Probability Density Function (PDF) in Fig.~\ref{fig:pdf_times}.
The plot shows a step function, and the width of each step is equal to $\unit[2.02]{s}$, 
which corresponds to the slotframe duration. 
Within each step, latency is uniformly distributed because periods at the application layer ($T_{\mathrm{update}}=\unit[30]{min}$) 
and at the MAC layer (that is, the repetition period of a cell in the slotframe, equal to $\unit[2.02]{s}$) 
are prime numbers and the two processes can be treated as they were independent.
The first (and higher) step is related to the frames that arrived to $\Nrx$ after exactly one transmission attempt, the second step refers to frames transmitted twice, and so on. 
The same plot also reports the Cumulative Distribution Function (CDF) of the swap latency $d_{\mathrm{SW}}$, which can be used to determine the expected number of \nuex{} that experienced a swap latency smaller than a given value.
For the same quantity $d_{\mathrm{SW}}$, the maximum latency is $d_{\mathrm{max}}=\unit[14.100]{s}$, which refers to a frame that was transmitted $7$ times (i.e., $\lceil \frac{d_{\mathrm{max}}}{2.02} \rceil$) before reaching the destination.

Regarding the double listening latency $d_{\mathrm{DL}}$, its average value is about $\unit[30]{s}$, which is equal to $T_{\mathrm{app}}$.
This is unsurprising, because only when a frame is transmitted in $C_{\mathrm{back}}$ 
(see arc \{5\} of the $\Nrx$ state machine in Fig.~\ref{fig:machines}), 
the new hopping function $\nu^n$ is definitely activated, and the double listening phase ends. 
Since packets are generated cyclically with period $T_{\mathrm{app}}=\unit[30]{s}$, excluding retransmissions, this frame typically arrives $\unit[30]{s}$ after the frame containing $\nu^n$. 
This is one of the drawbacks of \consp{}: in other words, the $d_{\mathrm{DL}}$ interval, 
in which a considerable amount of energy is wasted due to idle listening because both cells are active, depends on the link usage. 
However, in those links that experience higher traffic, that typically correspond to the levels in the tree network topology closest to the root node, where black listing techniques are more important to reduce the overall number of retransmissions, the interval between two successive packets is usually shorter than in the links close to the leaf nodes.

An important property is that, at least two frames are need to perform the whole \nuex{}. 
This is confirmed by the minimum value reported in the row of the table that refers to the total latency $d_{\mathrm{tot}}$.

\section{Conclusions}
\label{sec:conclusions}
In the context of black/white listing techniques, a crucial point is how to spread the information about the channels to be used for the transmission between nodes.
When the hopping sequence is defined on a per-link basis, only the two end points are involved.
In the case of inconsistency, communication in the network can be prevented, with possible definitive disconnections of subsets of nodes.

The \consp{} technique was proposed to counteract this problem by means of a backup cell. 
Each time a modification is triggered about the channels to be used for transmission over the link connecting two nodes, for a limited period of time both cells are exploited for communication. 
Doing so guarantees that the information seen by the two involved nodes is always updated in a coherent way, irrespective of the number of transmission errors that affected either data or acknowledgement frames.
The experimental analysis of \consp{}, performed by means of a simulator that was configured with data derived from a real setup, highlights its effectiveness. 
In particular, \consp{} does not affect communication latency, and has a small impact on energy consumption. 
Its main drawback is the need to reserve a backup cell per link, which can limit the number of nodes in dense networks.

Future works include improvements related to the \consp{} technique, which are aimed for instance to reduce energy consumption further, or to lower the number of backup cells that need to be reserved by the protocol. 
Future directions include the usage of \consp{} in the implementation of a black/white listing technique.

\bibliographystyle{IEEEtran}
\bibliography{bibliography}

\end{document}